\documentclass[journal]{IEEEtran}
\usepackage{cite}

\ifCLASSINFOpdf
\else
  \usepackage[dvips]{graphicx}
  \usepackage{subfigure}
\fi
\usepackage{amsmath}
\DeclareMathOperator{\sgn}{sgn}
\begin{document}
%
\title{A Multiscale Modeling of Triple-Heterojunction Tunneling FETs}
%
%
%

\author{Jun~Z.~Huang,
        Pengyu~Long,
        Michael~Povolotskyi,
        Hesameddin Ilatikhameneh,
        Tarek Ameen,
        Rajib Rahman,
        Mark~J.~W.~Rodwell, 
        and~Gerhard~Klimeck
\thanks{J. Z. Huang, P. Long, M. Povolotskyi, H. Ilatikhameneh, T. Ameen, R. Rahman, and G. Klimeck are with
the Network for Computational Nanotechnology and Birck Nanotechnology
Center, Purdue University, West Lafayette, IN 47907 USA (e-mail:
junhuang1021@gmail.com).}
\thanks{M. J. W. Rodwell is with the Department of Electrical and Computer
Engineering, University of California at Santa Barbara, Santa Barbara,
CA 93106-9560 USA.}
\thanks{Manuscript received xxx xx, 2017; revised xxxx xx, 2017. This work uses nanoHUB.org
computational resources operated by the Network for Computational
Nanotechnology funded by the U.S. National Science Foundation
under Grant EEC-0228390, Grant EEC-1227110, Grant EEC-0634750,
Grant OCI-0438246, Grant OCI-0832623, and Grant OCI-0721680. This
material is based upon work supported by the National Science Foundation
under awards 1509394, 1639958 and Semiconductor Research Corporation under award 2694.03.}}

\maketitle

\begin{abstract}
A high performance triple-heterojunction (3HJ) design has been previously proposed for tunneling FETs (TFETs). Compared with single heterojunction (HJ) TFETs, the 3HJ TFETs have both shorter tunneling distance and two transmission resonances that significantly improve the ON-state current ($I_{\rm{ON}}$). Coherent quantum transport simulation predicts, that $I_{\rm{ON}}=460\rm{\mu A/\mu m}$ can be achieved at gate length $Lg=15\rm{nm}$, supply voltage $V_{\rm{DD}}=0.3\rm{V}$, and OFF-state current $I_{\rm{OFF}}=1\rm{nA/\mu m}$. However, strong electron-phonon and electron-electron scattering in the heavily doped leads implies, that the 3HJ devices operate far from the ideal coherent limit. In this study, such scattering effects are assessed by a newly developed multiscale transport model, which combines the ballistic non-equilibrium Green's function method for the channel and the drift-diffusion scattering method for the leads. Simulation results show that the thermalizing scattering in the leads both degrades the 3HJ TFET's subthreshold swing through scattering induced leakage and reduces the turn-on current through the access resistance. Assuming bulk scattering rates and carrier mobilities, the $I_{\rm{ON}}$ is dropped from $460\rm{\mu A/\mu m}$ down to $254\rm{\mu A/\mu m}$, which is still much larger than the single HJ TFET case.
\end{abstract}


%
\IEEEpeerreviewmaketitle

\section{Introduction}
%
%
%
%
\IEEEPARstart{T}{unneling} field-effect transistor (TFET) is a steep subthreshold swing (SS) device that can operate at low supply voltage, so it is promising for low-power logic electronics application \cite{ionescu2011tunnel}. The low tunneling probability problem of TFETs can be mitigated by employing staggered- or broken- gap heterojunction (HJ) for the tunnel junction to reduce the tunnel barrier height and tunnel distance \cite{nayfeh2008,mohata2011demonstration,dey2013high}. However, even for broken-gap GaSb/InAs HJ TFETs, the tunnel probability is low due to quantum confinement induced band gap overlap~\cite{pala2015exploiting,long2016high,huang2017scalable}. A number of designs have been proposed to further improve the performance of the GaSb/InAs HJ TFETs~\cite{pala2015exploiting,long2016high,huang2017scalable,zhou2012novel,avci2013heterojunction,Carrillo2015InAs,Long2016drc,Long2016iprm,Huang2016p,Long2016a}. Among these designs, the triple HJ (3HJ) designs \cite{Long2016drc,Long2016iprm,Huang2016p,Long2016a} significantly boost the tunnel probability of the HJ TFETs by adding two additional, properly designed, HJs: one in the source and the other in the channel. For n-type 3HJ TFETs~\cite{Long2016drc}, atomistic quantum ballistic transport simulations have shown, that extremely high $I_{\rm{ON}}$ of $800\rm{\mu A/\mu m}$ ($460\rm{\mu A/\mu m}$) could be obtained at $Lg=30\rm{nm}$ ($15\rm{nm}$), $V_{\rm{DD}}=0.3\rm{V}$, and $I_{\rm{OFF}}=1\rm{nA/\mu m}$.

The 3HJ devices' promising ON/OFF current ratio is likely to be degraded by various scattering phenomena, as implied by the large spatial overlap of conduction- and valence-band local density of states (LDOS) at the OFF state~\cite{Long2016drc,Long2016iprm,Huang2016p,Long2016a}. The effect of electron-phonon scattering on the 3HJ design performance has been examined by the non-equilibrium Green's function (NEGF) approach in the self-consistent Born approximation (SCBA)~\cite{Long2016a}, considering only the diagonal components of the self-energy. It is found, that the electron-phonon scattering moderately degrades the transistor's I-V characteristics, but the $I_{\rm{ON}}$ at a given $I_{\rm{OFF}}$ and $V_{\rm{DD}}$ remains much larger than that of a GaSb/InAs HJ TFET, when simulated under equal intensity of phonon scattering~\cite{Long2016a}. The strong electron-electron scattering, such as the Auger generation \cite{Teherani2016Auger}, present in the heavily doped source (and drain) lead, however, has not been examined yet. Practically, this strong, close to thermalizing, effect of electron-electron scattering is very difficult to model in the NEGF approach with an explicit self-energy due to its non-local nature, as explained in more detail in Ref.~\cite{Long2016performance}. Therefore, we have developed an efficient empirical scattering model, using the NEMO5 tool~\cite{Steiger2011}. In this model, the central channel is treated as non-equilibrium and ballistic, while the leads are assumed to be in local thermal equilibrium with scattering rate derived from experimental carrier mobility. The local quasi-Fermi levels in the leads are determined by solving the drift-diffusion equations in them. Note, that the single scattering rate takes into account multiple scattering mechanisms present in the heavily doped leads, and the drift-diffusion solver allows us to include the lead serial resistance naturally.

This multiscale model, originally proposed in Ref.~\cite{Klimeck03} for resonant tunneling diode simulation, has been generalized to be able to simulate other structures, such as the ultra-thin-body (UTB) superlattice transistors~\cite{Long2016performance}. In this study, the model is further extended to consider both electron and hole transport, required for a band-to-band tunneling device simulation. This model is then employed to systematically analyze the 3HJ TFETs by varying the scattering strength, and corresponding mobility, of the leads. The device behaviors under channel length scaling are studied and, finally, the device performances are benchmarked against conventional HJ TFETs.

\begin{figure}\centering
\includegraphics[width=8.7cm]{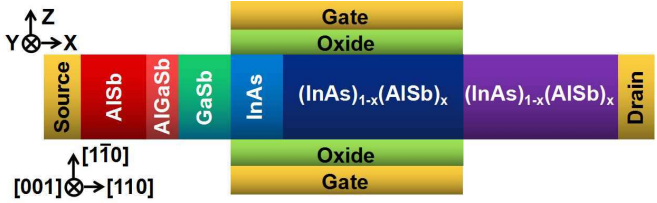}
\caption{\label{fig:device_geo} Device structure and material compositions of the n-type 3HJ TFET.
The p-type doped source lead consists of a 5.6nm AlSb layer with doping density $N_A=3\times10^{19}/\rm{cm}^3$,
a 1.2nm $\rm{Al}_{0.5}\rm{Ga}_{0.5}\rm{Sb}$ grade layer and a 3.3nm GaSb layer, both with $N_A=5\times10^{19}/\rm{cm}^3$.
The intrinsic channel consists of a 3nm InAs layer and a 12nm $\rm{(InAs)}_{0.79}\rm{(AlSb)}_{0.21}$ layer.
The n-type doped drain lead consists of a 10nm $\rm{(InAs)}_{0.79}\rm{(AlSb)}_{0.21}$ layer with doping density $N_D=2\times10^{19}/\rm{cm}^3$.
The body thickness and the oxide thickness are both 1.8nm with oxide dielectric constant $\epsilon_r=9$. The confinement (transport) crystal orientation is [1$\bar{1}$0] ([110]).}
\end{figure}
\begin{figure}\centering
\includegraphics[width=8.7cm]{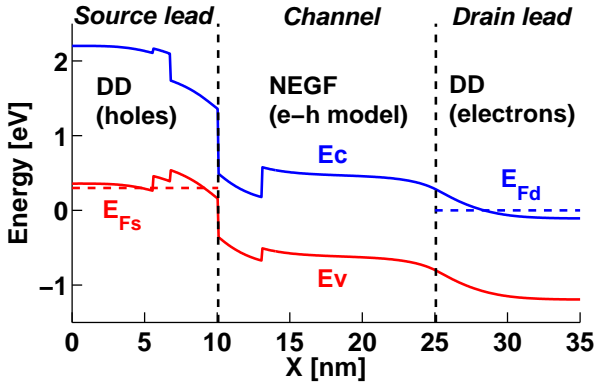}
\caption{\label{fig:region} Region boundaries of the multiscale model. The OFF-state band diagram and the Fermi levels are superimposed.
We couple the drift-diffusion equations for the holes (electrons) in the source (drain) lead and the NEGF equations for both the electrons and the holes in the channel.}
\end{figure}

\section{The Triple-Heterojunction Design}
We consider n-type 3HJ TFETs with double-gate UTB structures~\cite{Long2016drc}. The device geometry and the optimized design parameters
are shown in Fig.~\ref{fig:device_geo}. The added HJ in the channel is InAs/$\rm{(InAs)}_{0.79}\rm{(AlSb)}_{0.21}$, while the added HJ in the source is AlSb/GaSb with an $\rm{Al}_{0.5}\rm{Ga}_{0.5}\rm{Sb}$ grading layer. The materials are chosen so, that the channel material ($\rm{(InAs)}_x\rm{(AlSb)}_{1-x}$ in general) has a higher conduction band edge, than in the InAs layer, while the source material ($\rm{Al}_y\rm{Ga}_{1-y}\rm{Sb}$ in general) has a lower valence band edge, than in the GaSb layer. In addition, these materials have very close lattice constants around 6.1{\AA} \cite{Kroemer2004}, hence no significant strain is induced. As shown in Refs.~\cite{Long2016drc,Huang2016p}, the band offsets greatly enhance the electric field at the GaSb/InAs tunnel junction and create two resonant states in the GaSb and InAs quantum wells, improving the tunnel probability at ON state. The larger band gap and transport effective mass of the $\rm{(InAs)}_x\rm{(AlSb)}_{1-x}$ channel, compared with the original InAs channel, also reduce the ambipolar and the source-to-drain tunneling leakage in the sub-threshold region. To maximize the electric field at the tunnel junction, the band offsets of the added two HJs should be as large as possible and they should also balance each other, leading to optimal mole fractions $x=0.79$ and $y=1$. The GaSb and InAs quantum well widths are then adjusted to have the resonant state levels closely aligned in the Fermi conduction window at the ON state.

\section{The Multiscale Simulation Approach}
The device is divided into three regions, as shown in Fig.~\ref{fig:region}. The assumption is, that in the doped source and drain regions
the strong scattering drives the free charge carriers into local thermal equilibrium. Therefore, we solve two drift-diffusion (DD) current equations for the thermalized holes in the source and the thermalized electrons in the drain, respectively. Since quantum confinements are critical in the leads, we obtain the hole and electron density of states from the Green's functions.
The minority carrier current in the leads is neglected, because both the source and the drain leads are heavily doped. In the intrinsic channel region, the scattering rate is much lower, so the quantum-mechanical band-to-band tunneling is the major transport mechanism. Hence, in the channel region we solve the ballistic NEGF equations involving both the electrons and the holes.
The contact self energies and quasi-Fermi levels for the quantum domain boundaries are taken from the lead surface Green's functions and DD equations, respectively. The central quantum domain provides ballistic current, that is used as a boundary condition for the DD equations in the leads. The DD equations in the leads require LDOS, that is computed from Green's function in a recursive way, using the quantum domain Green's function as a boundary condition.

\subsection{Numerical Model Detail}
The device Hamiltonian is modeled using  $sp^3d^5s^*$ tight binding (TB) basis with spin-orbital interaction.
The parameters are taken from Ref.~\cite{Tan2016}. The alloys are described using the virtual crystal approximation (VCA)
with their TB parameters linearly interpolated from their corresponding binaries. In order to utilize the efficient recursive Green's function (RGF) algorithm~\cite{lake1997single}, the whole device is partitioned into a set of layers. The layers are numbered with $l=1,2,\cdots,P$ for the layers in the source lead, $l=P+1,P+2,\cdots,Q-1$ for the layers in the channel, and $l=Q,Q+1,\cdots,N$ for the layers in the drain lead.

We solve ballistic NEGF equations in the central channel (between layer $P$ and layer $Q$), and the current passing through it $I_{PQ}$ is computed by the transmission $T_{PQ}$ weighted by the Fermi function difference:
\begin{align}
I_{PQ}&=\frac{q}{h}\int\frac{dk_y}{2\pi}\int\frac{dE}{2\pi}\times \nonumber \\
&T_{PQ}\left(k_y,E\right)\left[f\left(E-E_{F,P}\right)-f\left(E-E_{F,Q}\right)\right],
\end{align}
where $q$ is the electron charge and $h$ is the Planck constant. The integration is performed over the transverse momentum $k_y$ and energy $E$.

Considering both electrons and holes, the charge density in the central channel per orbital~$\alpha$ in layer~$l$ can be calculated as~\cite{Guo2004Toward}:
\begin{align}
\rho_{l,\alpha}=&\int\frac{dk_y}{2\pi}\int\frac{dE}{2\pi}\sgn\left(E-E_{N,l}\right)\times \nonumber \\
&\{A^L_{l,\alpha}\left(k_y,E\right)f\left[\sgn\left(E-E_{N,l}\right)\left(E-E_{F,P}\right)\right]\nonumber\\
&+A^R_{l,\alpha}\left(k_y,E\right)f\left[\sgn\left(E-E_{N,l}\right)\left(E-E_{F,Q}\right)\right]\},
\end{align}
where $A^L_{l,\alpha}$ and $A^R_{l,\alpha}$ are the diagonal components of the left- and right-connected spectral functions in the channel, and $\sgn\left(E\right)$ is the sign function. $E_{N,l}$ is the layer dependent threshold (charge neutrality level) defined as the middle of the band gap:
\begin{equation}
E_{N,l}=0.5\left(E_{V,l}+E_{C,l}\right),
\end{equation}
where $E_{V,l}$ and $E_{C,l}$ are the layer-dependent valence and conduction band edge, respectively.

We solve DD equations for holes in the source lead and for electrons in the drain lead:
\begin{align}
J^h & =\mu^h p\nabla E_F, \\
J^e & =\mu^e n\nabla E_F,
\end{align}
where, $\mu^h$ and $\mu^e$ are the hole and electron mobilities, $p$ and $n$ are the hole and electron densities, and $E_F$ is the major carrier quasi-Fermi level.
For simplicity we assume, that the Fermi level is constant over a single layer. Therefore, we have to determine a set of layer-dependent quasi-Fermi levels $E_{F,l}$ ($l=1,2,\cdots,P,Q,Q+1,\cdots,N$).

For the hole transport in the source lead, current conservation in the transport direction requires that
\begin{align}
\mu^h p_l\nabla E_{F,l}-J_{PQ} & \approx \mu^h p_l \frac{E_{F,l}-E_{F,l-1}}{\Delta x}-\frac{I_{PQ}}{T_z} = 0,~\nonumber\\
&{\rm{for}}~l=1,2,\cdots,P ,
\end{align}
where the backward difference is used to discretize the differential operator $\nabla$. $\Delta x$ and $T_z$ are the layer thickness in the $x$ and $z$ direction, respectively. Similarly, for the electron transport in the drain lead:
\begin{align}
\mu^e n_l\nabla E_{F,l}-J_{PQ} & \approx \mu^e n_l \frac{E_{F,l+1}-E_{F,l}}{\Delta x}-\frac{I_{PQ}}{T_z} = 0,~\nonumber\\
&{\rm{for}}~l=Q,Q+1,\cdots,N ,
\end{align}
where the forward difference is used to discretize the differential operator $\nabla$. $p_l$ and $n_l$ are averaged over layer $l$.

The hole and electron orbital resolved density in the leads is computed by integrating the diagonals of the lead spectral function $A_{l,\alpha}$ weighted by either hole or electron Fermi-Dirac function with local quasi-Fermi level $E_{F,l}$:
\begin{multline}
p_{l,\alpha}=\int\frac{dk_y}{2\pi}\int\frac{dE}{2\pi} \bigl( \frac{1}{2}\left[-\sgn\left(E-E_{N,l}\right)+1\right] \times \\
A_{l,\alpha}\left(k_y,E\right)f\left(E_{F,l}-E\right) \bigr),~{\rm{for}}~l=1,2,\cdots,P.
\end{multline}
\begin{multline}
n_{l,\alpha}=\int\frac{dk_y}{2\pi}\int\frac{dE}{2\pi}\bigl( \frac{1}{2}\left[\sgn\left(E-E_{N,l}\right)+1\right] \times \\
A_{l,\alpha}\left(k_y,E\right)f\left(E-E_{F,l}\right) \bigr),~{\rm{for}}~l=Q,Q+1,\cdots,N.
\end{multline}

To account for a broadening due to scattering in the leads, a small imaginary potential $-i\eta$ is added to the diagonal of the lead Hamiltonian \cite{Klimeck95}.
The $\eta$ is constant below the confined valence band edge and above the confined conduction band edge, while it is exponentially decaying in the band gap:
\begin{equation}
\eta_l(E) = \begin{cases}
         \eta_h,&E < E_{V,l}\\
         \eta_h\exp\left(\frac{E_{V,l}-E}{E_0}\right),&E_{V,l} < E < \frac{E_{C,l}+E_{V,l}}{2}\\
         \eta_e\exp\left(\frac{E-E_{C,l}}{E_0}\right),&\frac{E_{C,l}+E_{V,l}}{2}<E<E_{C,l}\\
         \eta_e,&E_{C,l} < E,
        \end{cases}
\end{equation}
where $\eta_e$ and $\eta_h$ are the electron and hole scattering constants (rates), and decaying constant $E_0$ is the Urbach parameter.
The $\eta_h$ here is related to hole momentum relaxation
time $\tau_h=\hbar/2\eta_h$ and hole mobility $\mu^h=q\tau_h/m_h^*$, where $m_h^*$~is the hole effective mass;
similar expressions relate $\eta_e$ to electron momentum relaxation time $\tau_e$ and electron mobility $\mu^e$.

%

\begin{figure}\centering
\includegraphics[width=8cm]{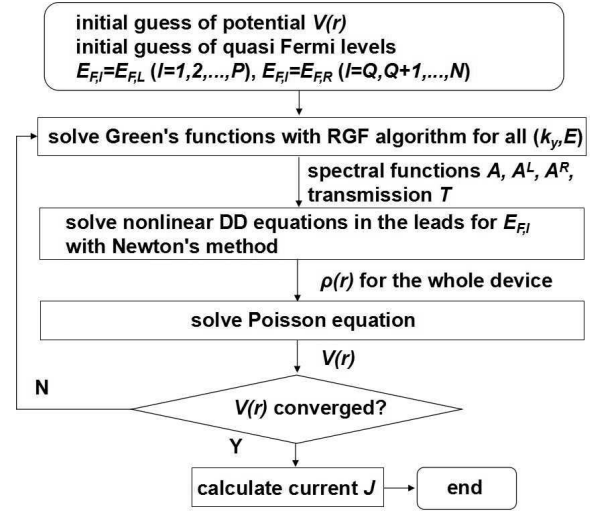}
\caption{\label{fig:flow_chart} The self-consistent simulation flow.}
\end{figure}

\subsection{Program Flow}
Given an electric potential distribution of the device $V\left(\mathbf{r}\right)$, the special RGF algorithm in Refs.~\cite{lake1997single,Klimeck95} is employed
to compute efficiently the transmission $T_{PQ}$, the $A_{l,\alpha}$ in the leads, as well as the $A^L_{l,\alpha}$ and $A^R_{l,\alpha}$ in the channel. Here, the RGF calculations
for different transverse momentum $k_y$ and energy $E$ are distributed over different CPU cores. Then, non-linear equations (6) and (7),
together with boundary conditions $E_{F,0}=E_{F,L}$ and $E_{F,N+1}=E_{F,R}$ ($E_{F,L}$ and $E_{F,R}$ are the Fermi levels of the left and right contacts),
are solved iteratively using the Newton method to find $E_{F,l}$. Note that the Jacobian matrix is sparse and can be calculated analytically
(derivative of the Fermi-Dirac function). With quasi-Fermi levels $E_{F,l}$ determined, the charge density of the whole device is updated using (2), (8) and (9),
considering both electrons and holes in the channel (see Eq.~2) and neglecting minority carriers in the leads.
The non-linear Poisson equation is then solved to find a new $V\left(\mathbf{r}\right)$. The self-consistent simulation flow is summarized in Fig.~\ref{fig:flow_chart}. For the device dimensions shown in Fig.~\ref{fig:device_geo} and for one bias point, it usually takes around one hour to get a converged solution on 96 CPU cores (6 dual 8-core Intel Xeon-E5 CPUs), which is roughly four to five times slower than ballistic simulations with quantum transmitting boundary method~\cite{Luisier2006}.

\section{Simulation Results}
We use $\eta_h=6{\rm meV}$ for the holes at the GaSb source,
which translates to $\tau_h=55{\rm fs}$, and $\mu^h=241\rm{cm^2/\left(V\cdot s\right)}$ assuming bulk heavy
hole effective mass $m^*_{hh}=0.4m_0$. This mobility value is comparable to the experimental hole mobility of
bulk GaSb at $5\times10^{19}{\rm cm^{-3}}$ doping density, which is around $250\rm{cm^2/\left(V\cdot s\right)}$~\cite{Mart2004}. We use $\eta_e=9{\rm meV}$ for the electrons at the $\rm{(InAs)}_{0.79}\rm{(AlSb)}_{0.21}$ drain, which translates to $\tau_e=36.6{\rm fs}$,
and $\mu^e=1497\rm{cm^2/\left(V\cdot s\right)}$ using bulk electron effective mass $m^*_e=0.043m_0$, obtained
from TB calculation using VCA. This mobility value is comparable to the
experimental electron mobility of bulk $\rm{In}_{0.53}\rm{Ga}_{0.47}\rm{As}$ at $2\times10^{19}{\rm cm^{-3}}$ doping density,
which is $1,422\rm{cm^2/\left(V\cdot s\right)}$~\cite{Sotoodeh2000}. We use the experimental mobility value
of $\rm{In}_{0.53}\rm{Ga}_{0.47}\rm{As}$ as a reference, because there is no experimental mobility data
for $\rm{(InAs)}_{0.79}\rm{(AlSb)}_{0.21}$ to the best of our knowledge, and the electron effective
mass of $\rm{(InAs)}_{0.79}\rm{(AlSb)}_{0.21}$ is very close to that of $\rm{In}_{0.53}\rm{Ga}_{0.47}\rm{As}$ which is $0.041m_0$.
Quantum confinement may change the carrier mobility and scattering rate,
that needs to be assessed in future studies.
We set the Urbach parameter $E_0$ to be the thermal energy at room temperature, {\it i.e.}, 26meV, which has been justified and
used in Ref.~\cite{Khayera2011} for studying the band tail effects in TFETs.

\begin{figure}\centering
\includegraphics[width=8.7cm]{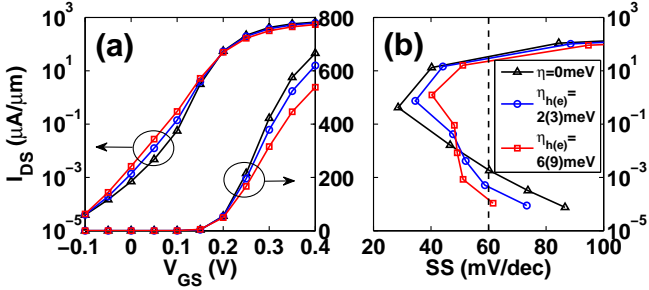}
\caption{\label{fig:Id_Vg_scattering_vs_ballistic} (a) $I_{\rm{DS}}$-$V_{\rm{GS}}$ characteristics (at $V_{\rm{DS}}=0.3\rm{V}$) in both logarithmic scale and linear scale. Different $\eta$ are compared. The $\eta$ are associated with $\mu$. $\eta=0$ corresponds
to $\mu=\infty$ for both leads. $\eta_h=2\rm{meV}$ and $\eta_e=3\rm{meV}$ correspond to $\mu^h=724\rm{cm^2/\left(V\cdot s\right)}$ and
$\mu^e=4492\rm{cm^2/\left(V\cdot s\right)}$ for the source and drain leads, respectively. $\eta_h=6\rm{meV}$ and $\eta_e=9\rm{meV}$ correspond
to $\mu^h=241\rm{cm^2/\left(V\cdot s\right)}$ and $\mu^e=1497\rm{cm^2/\left(V\cdot s\right)}$ for the source and drain leads, respectively. (b) $I_{\rm{DS}}$-SS extracted from the curves in (a).}
\end{figure}

\begin{figure}\centering
\includegraphics[width=8.7cm]{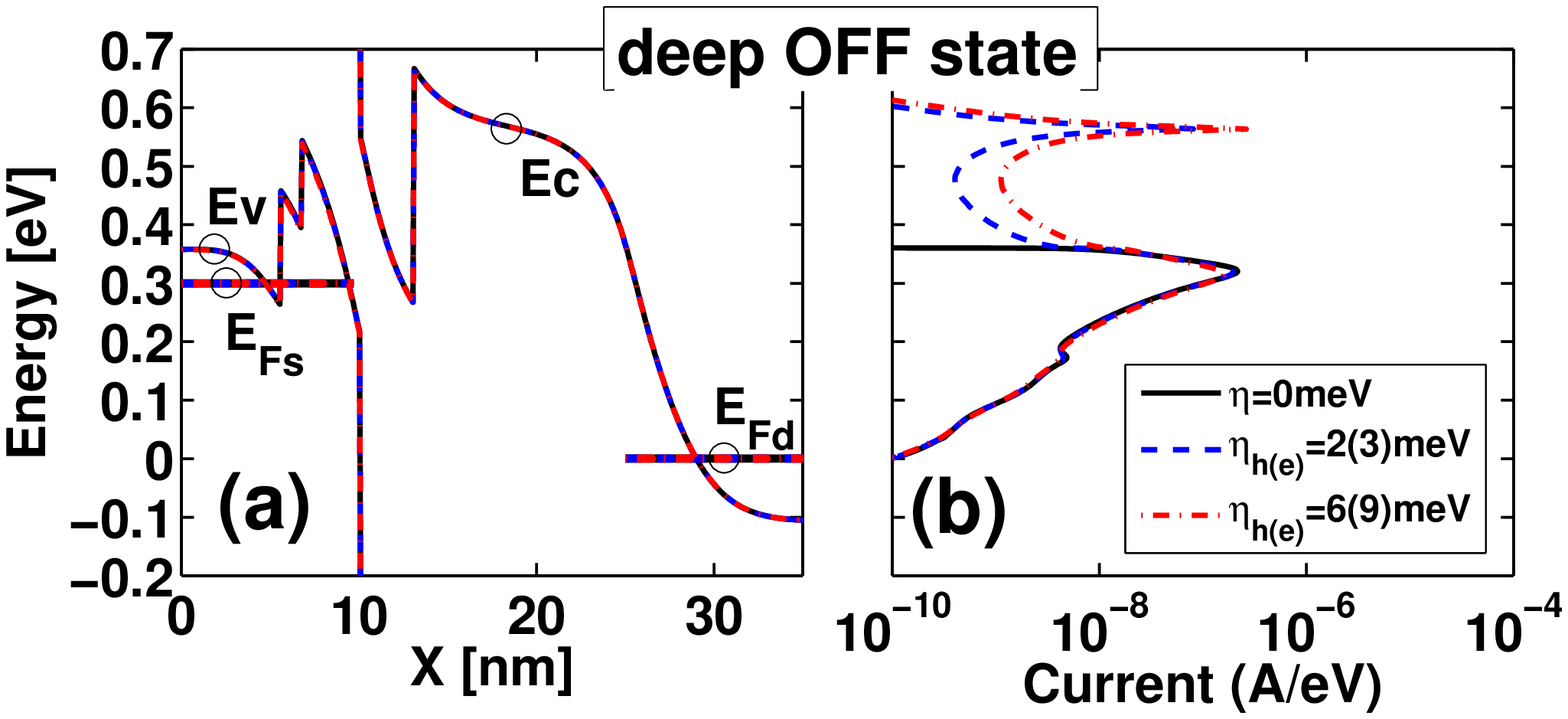}
\includegraphics[width=8.7cm]{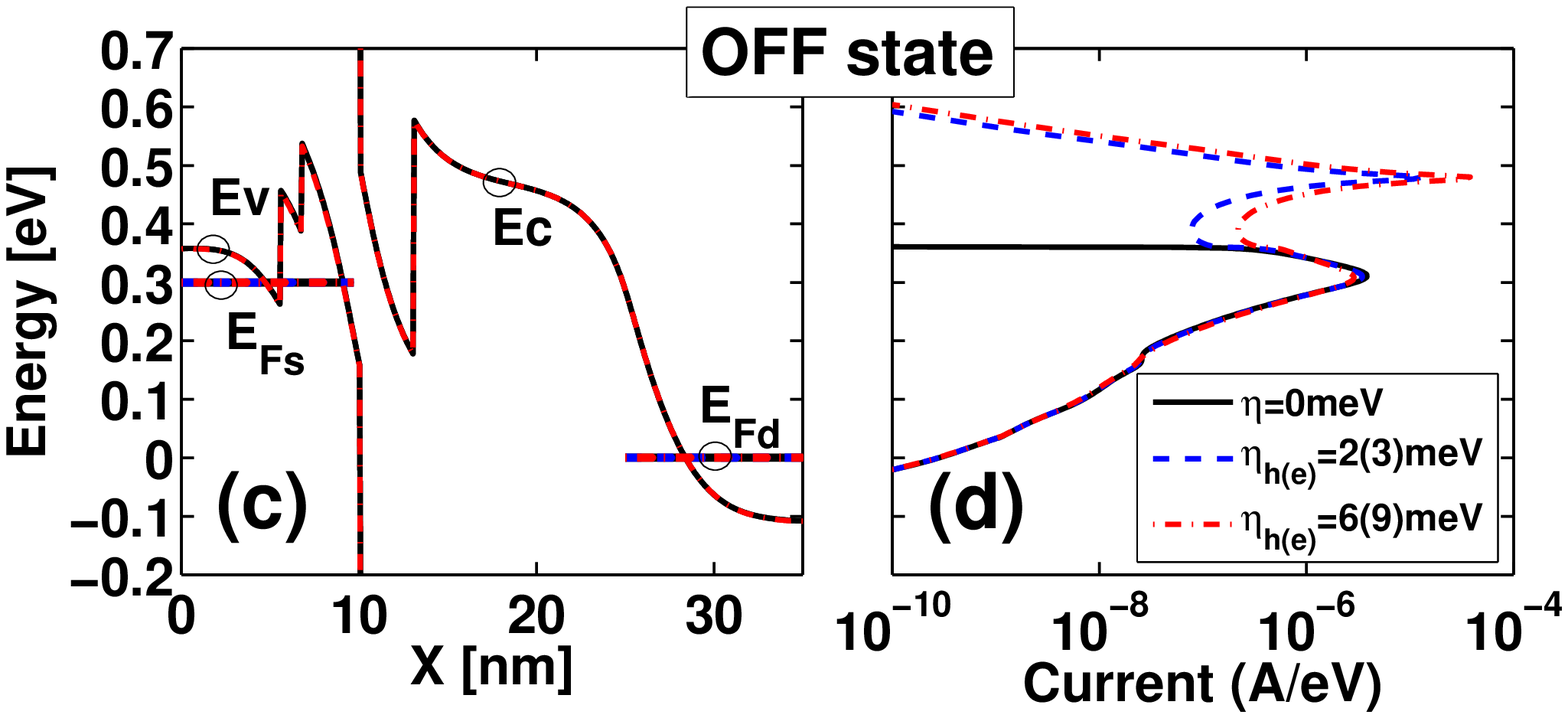}
\includegraphics[width=8.7cm]{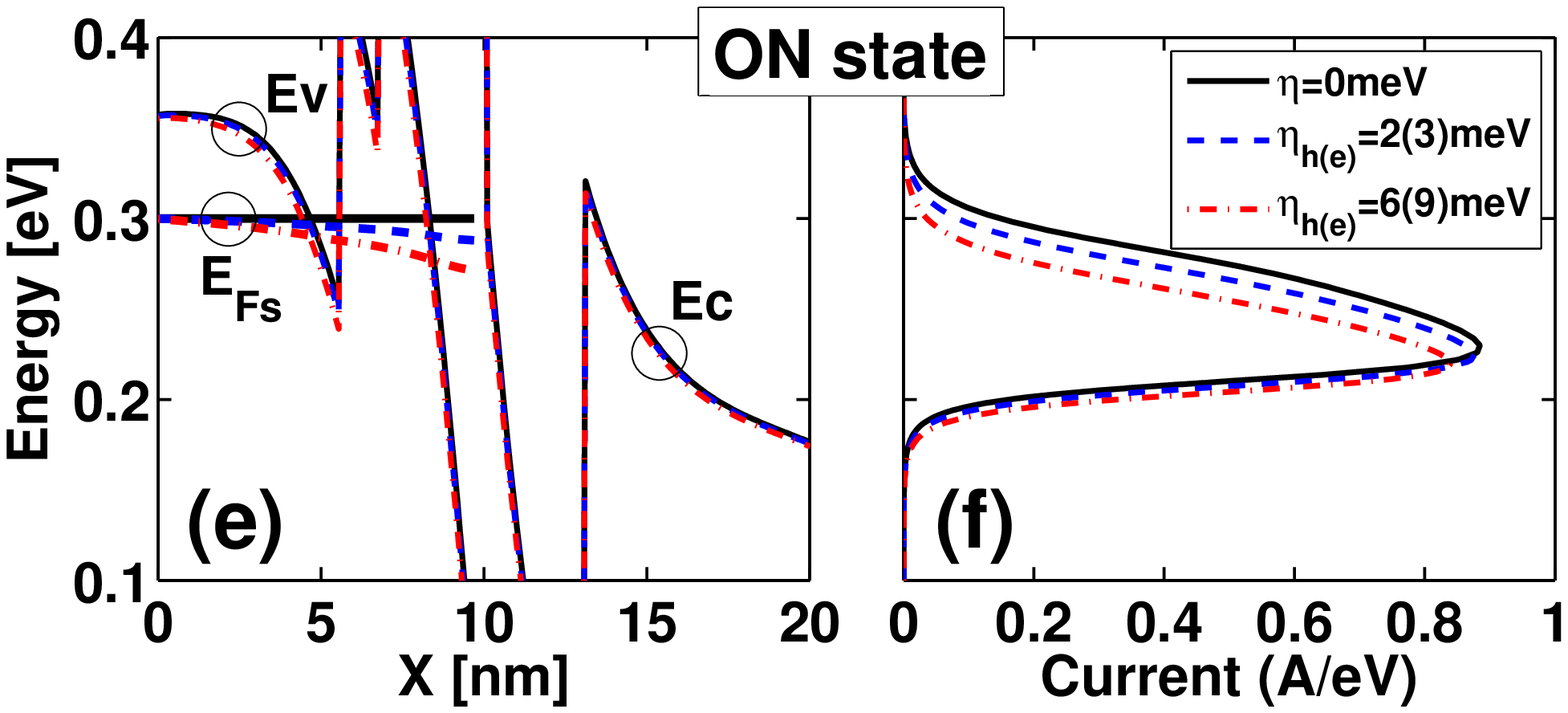}
\caption{\label{fig:pot_trans_scattering_vs_ballistic} Band diagrams ($V_{\rm{DS}}=0.3\rm{V}$) and current spectra ($k_y=0$) at $V_{\rm{GS}}=-0.1\rm{V}$ (a and b),
$V_{\rm{GS}}=0\rm{V}$ (c and d), and $V_{\rm{GS}}=0.3\rm{V}$ (e and f). Different $\eta$ and corresponding $\mu$ are compared.
In (a), (c), and (e), the quasi-Fermi levels in the leads are also superimposed.}
\end{figure}

The transfer characteristics for different scattering rates, and, hence, mobilities, are simulated and plotted in Fig.~\ref{fig:Id_Vg_scattering_vs_ballistic} (a). The extracted SS is plotted in Fig.~\ref{fig:Id_Vg_scattering_vs_ballistic} (b).
They show that both the SS and the turn-ON current change in the presence of scattering. With fixed $I_{\rm{OFF}}=1\rm{nA/\mu m}$ and at $V_{\rm{DD}}=0.3\rm{V}$, the $I_{\rm{ON}}$ is $460\rm{\mu A/\mu m}$ for $\eta=0meV$. It reduces by 23\% to $352\rm{\mu A/\mu m}$ for $\eta_{h(e)}=2(3)meV$ and by 45\% to $254\rm{\mu A/\mu m}$ for $\eta_{h(e)}=6(9)meV$. In the subthreshold region, as explained in Fig.~\ref{fig:pot_trans_scattering_vs_ballistic} (a)-(d), the scattering in the leads induces an additional peak in the current spectra,
which becomes bigger, as the scattering rate increases. At the deep OFF state ($V_{\rm{GS}}=-0.1\rm{V}$), as shown in
Fig.~\ref{fig:pot_trans_scattering_vs_ballistic} (a) and (b), the direct source-to-drain tunneling dominates, and, therefore, the total current does not
increase significantly. At the OFF state ($V_{\rm{GS}}=0\rm{V}$), as shown in Fig.~\ref{fig:pot_trans_scattering_vs_ballistic} (c) and (d), the
scattering leakage dominates, and, therefore, the total current increases significantly. At the ON state ($V_{\rm{GS}}=0.3\rm{V}$), as explained in
Fig.~\ref{fig:pot_trans_scattering_vs_ballistic} (e) and (f), the source serial resistance induces notable potential and quasi-Fermi level drops in the source lead,
which become more pronounced as the mobility decreases. The potential drop reduces the transmission and the quasi-Fermi level drop reduces the Fermi
conduction window, both decreasing the total current passing through the central channel. Note, that attempt to reduce the source serial resistance by increasing the doping density would result in large source Fermi degeneracy that in turn degrades the SS. The potential and quasi-Fermi level drop in the drain lead are less pronounced (not shown here) due to the smaller drain serial resistance as a result of the higher electron mobility. To quantify the most relevant mechanism among these two, {\it i.e.}, the scattering induced leakage and the serial resistance, for the $I_{\rm{ON}}$ reduction, we simulated $\mu=\infty$ (keeping $\eta_{h(e)}=6(9)\rm{meV}$) and $\eta=0$ (keeping $\mu^{h(e)}=241(1497)\rm{cm^2/\left(V\cdot s\right)}$) separately, and found that the $I_{\rm{ON}}$ (at $V_{\rm{DD}}=0.3\rm{V}$ and $I_{\rm{OFF}}=1\rm{nA/\mu m}$) drops by 29\% due to the scattering-induced leakage and by 24\% due to the serial resistance. Therefore, the two mechanisms have similar effects on the device $I_{\rm{ON}}/I_{\rm{OFF}}$ ratio.

\begin{figure}\centering
\includegraphics[width=8.7cm]{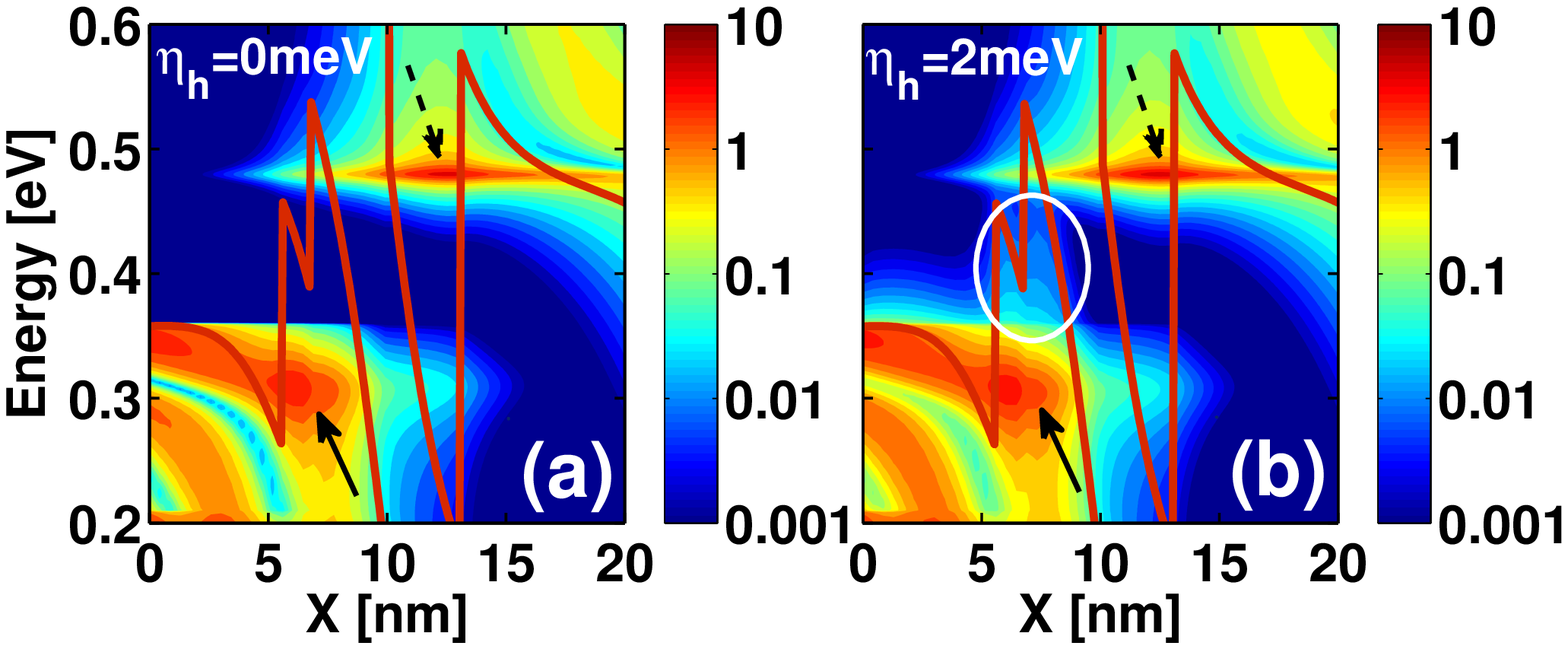}
\includegraphics[width=8.7cm]{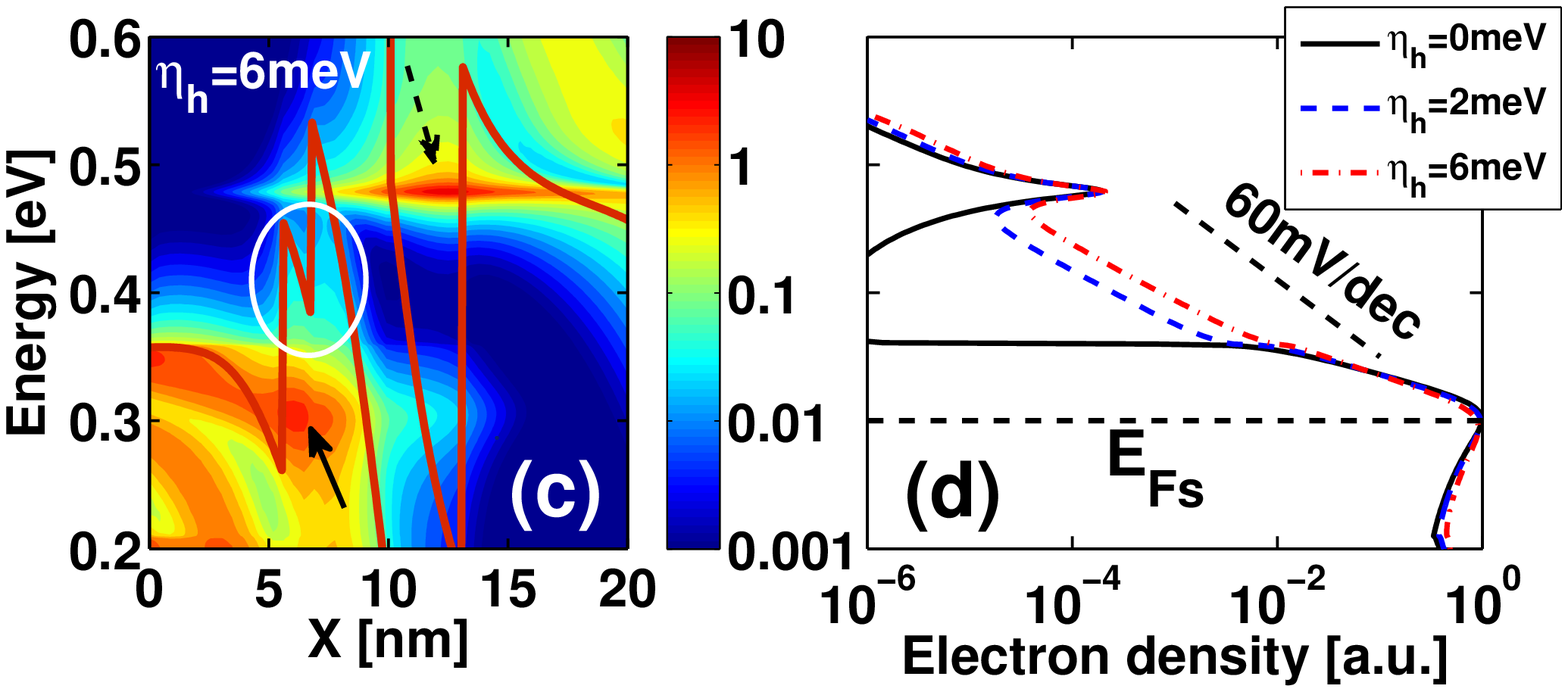}
\caption{\label{fig:ldos} logarithmic scale LDOS with the superimposed band diagram at $k_y=0$, $V_{\rm{DS}}=0.3\rm{V}$, and $V_{\rm{GS}}=0\rm{V}$, for $\eta_h=0$ (a), $\eta_h=2\rm{meV}$ (b), and $\eta_h=6\rm{meV}$ (c). The source (channel) resonant state is marked with solid (dashed) arrows. The scattering induced LDOS in the source notch is marked with white circles. (d) The electron spectral density at $\left(x,~y,~z\right)=\left(6.9,~0.46,~1.1\right)\rm{nm}$.}
\end{figure}

To explain the additional spectral current peak in the subthreshold region when scattering is present, we compare the LDOS at the OFF state for different scattering rates. For $\eta_h=0$, essentially ballistic, as shown in Fig.~\ref{fig:ldos} (a), the channel resonant state cannot connect to the source contact due to the lack of states in the source band gap. Therefore, the only leakage path is the direct source-to-drain tunneling leakage, peaked at the source resonant state energy level. While for $\eta_h=2{\rm meV}$ and $\eta_h=6{\rm meV}$, as shown in Fig.~\ref{fig:ldos} (b) and (c), the source valance band states, in particular the source resonant state, are broadened, introducing states into the source band gap and into the source potential notch. As a result, the channel resonant state is connected to the source, giving rise to the additional leakage path. It is worth mentioning that inelastic scattering mechanisms, such as the optical phonon scattering and the electron-electron scattering, are captured here, because in our model the $-i\eta_h$ is also added to the band gap (through the exponentially decaying function) and to the source potential notch where the ballistic LDOS is zero. The electron spectral densities for the three cases, sampled in the source quantum well, are compared in Fig.~\ref{fig:ldos} (d). It is shown, that the slope of the spectral density increases as the scattering rate increases, and with bulk scattering rate $\eta_h=6{\rm meV}$ the slope is still below 60mV/dec thermal limit. This explains why the SS of the scattering I-V curves, shown in Fig.~\ref{fig:Id_Vg_scattering_vs_ballistic} (b), is still less than 60mV/dec (for $10^{-3}\rm{\mu A/\mu m}<I_{\rm{DS}}<10\rm{\mu A/\mu m}$). For 1-D broken-gap TFETs featuring a similar potential notch in the source~\cite{koswatta2010possibility}, in the presence of electron-phonon scattering, sub-60mV/dec SS was also obtained, which was ascribed to the suppression of density of states (DOS) in the notch {\it and} nonequilibrium carrier distribution below the thermionic limit in the notch. In our simulation, the reduced DOS in the source notch solely accounts for the sub-60mV/dec SS.

\begin{figure}\centering
\includegraphics[width=8.7cm]{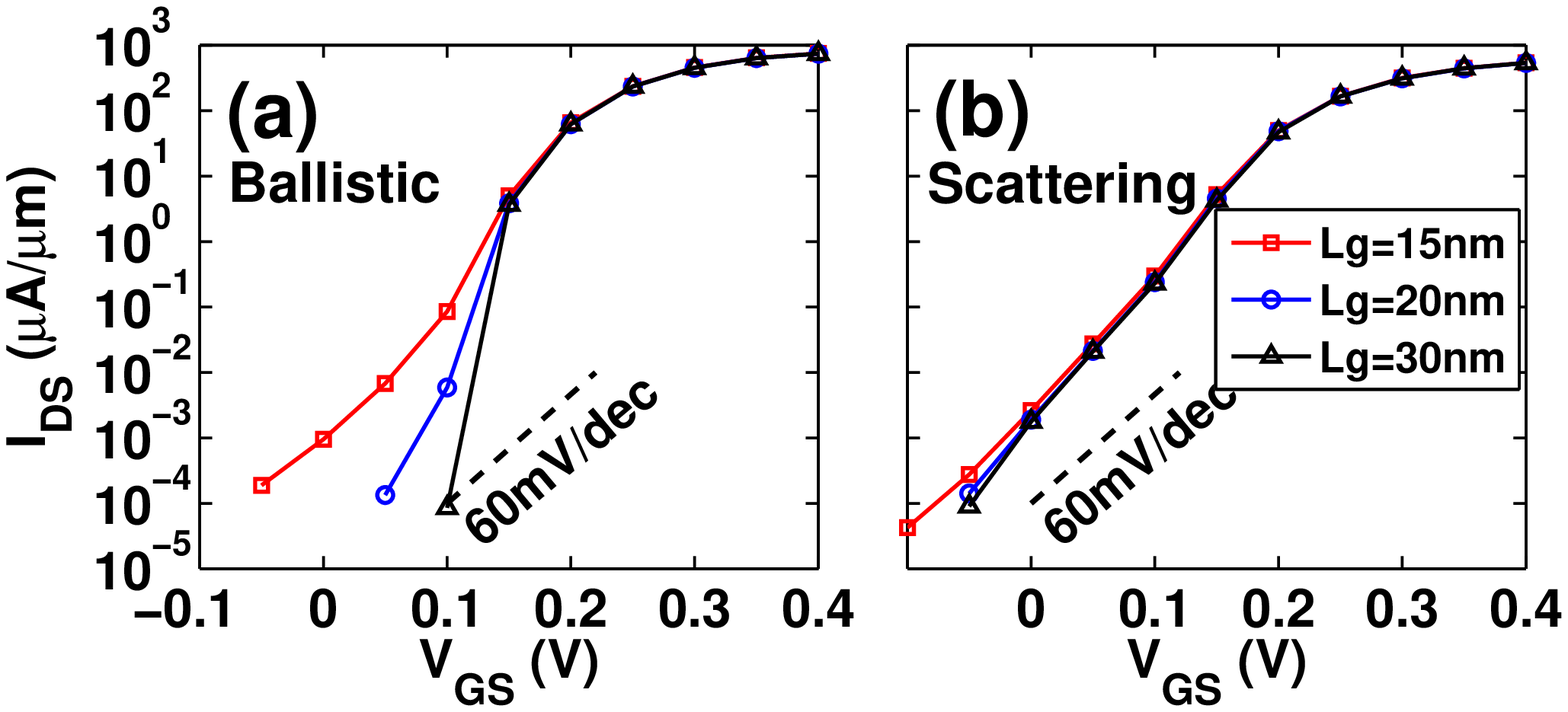}
\caption{\label{fig:Id_Vg_scaling} $I_{\rm{DS}}$-$V_{\rm{GS}}$ (at $V_{\rm{DS}}=0.3\rm{V}$) for different channel length Lg=15nm, Lg=20nm, and Lg=30nm. (a) Ballistic simulations, (b) scattering simulations with $\eta_h=6\rm{meV}$ ($\eta_e=9\rm{meV}$) and $\mu^h=241\rm{cm^2/\left(V\cdot s\right)}$ ($\mu^e=1497\rm{cm^2/\left(V\cdot s\right)}$) for the source (drain) lead.}
\end{figure}

\begin{figure}\centering
\includegraphics[width=8.7cm]{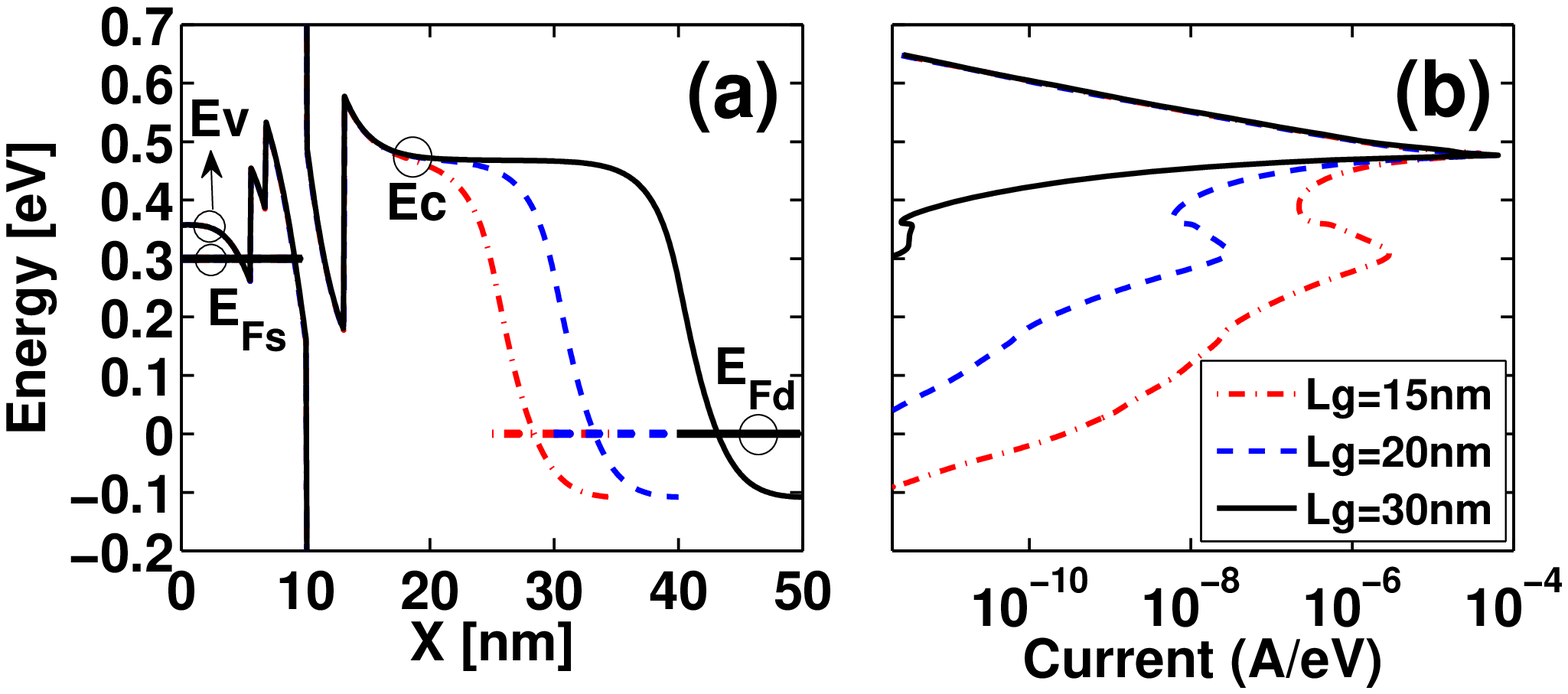}
\caption{\label{fig:transmission_scale} Band diagram (a) and current spectra (b) at $k_y=0$, $V_{\rm{DS}}=0.3\rm{V}$
and $V_{\rm{GS}}=0$ for channel length Lg=15nm, Lg=20nm, and Lg=30nm. For all
curves, $\eta_h=6\rm{meV}$ ($\eta_e=9\rm{meV}$) and $\mu^h=241\rm{cm^2/\left(V\cdot s\right)}$ ($\mu^e=1497\rm{cm^2/\left(V\cdot s\right)}$) are used
for the source (drain) lead.}
\end{figure}

We found that the channel length scaling behaviors are very different when lead scattering is included, as compared in Fig.~\ref{fig:Id_Vg_scaling}. For the ballistic case, the SS improves as the channel length (Lg) increases, which is easily understood because longer Lg suppresses the source-to-drain tunneling leakage. For the scattering case, when Lg is increased from 15nm to 20nm, the SS is slightly improved, but when Lg is further increased to 30nm, the SS is not improved further. The explanations are shown in Fig.~\ref{fig:transmission_scale} for the scattering case, where the band diagrams and current spectra at different Lg are compared. As can be seen, there are two peaks in the spectra, the lower one corresponds to direct source-to-drain tunneling through the source resonant state and the upper one
corresponds to the scattering induced leakage through the channel resonant state. As Lg increases, the direct tunneling leakage is
gradually suppressed but the scattering induced leakage remains unchanged. At Lg=15nm, the direct tunneling leakage is significant and,
therefore, increasing Lg helps reduce the total leakage. At Lg=20nm, the scattering induced leakage already dominates and, consequently, increasing Lg further does not help reduce the total leakage further.


\begin{figure}\centering
\includegraphics[width=8.7cm]{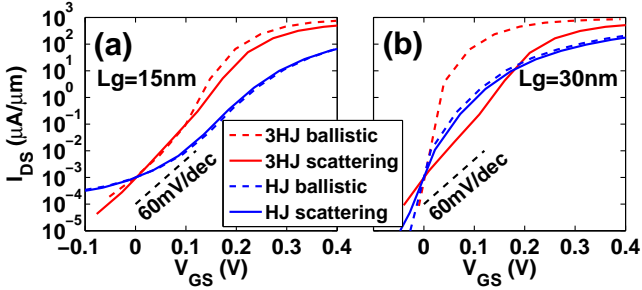}
\caption{\label{fig:Id_Vg_comparison} Comparison of $I_{\rm{DS}}$-$V_{\rm{GS}}$ (at $V_{\rm{DS}}=0.3\rm{V}$) between the 3HJ TFET and
the HJ TFET for channel length Lg=15nm (a) and Lg=30nm (b). For all scattering curves, $\eta_h=6\rm{meV}$ ($\eta_e=9\rm{meV}$)
and $\mu^h=241\rm{cm^2/\left(V\cdot s\right)}$ ($\mu^e=1497\rm{cm^2/\left(V\cdot s\right)}$) are used for the source (drain) lead.
The threshold voltages are all adjusted for the same $I_{\rm{OFF}}=1\rm{nA/\mu m}$.}
\end{figure}

To benchmark the ultimate device performances, we compare $I_{\rm{ON}}$ of the 3HJ TFETs with the HJ TFETs at $V_{\rm{DD}}=0.3\rm{V}$
and $I_{\rm{OFF}}=1\rm{nA/\mu m}$, for a short channel length case (Lg=15nm) and a long channel length case (Lg=30nm), as shown in Fig.~\ref{fig:Id_Vg_comparison}.
At Lg=15nm, the scattering reduces $I_{\rm{ON}}$ of the 3HJ TFET from $460\rm{\mu A/\mu m}$ to $254\rm{\mu A/\mu m}$, while it does not change $I_{\rm{ON}}$ of the
HJ TFET that much, which is $13\rm{\mu A/\mu m}$. At Lg=30nm, the scattering reduces $I_{\rm{ON}}$ of the 3HJ TFET from $767\rm{\mu A/\mu m}$ down to
$280\rm{\mu A/\mu m}$ and reduces $I_{\rm{ON}}$ of the HJ TFET from $87\rm{\mu A/\mu m}$ down to $75\rm{\mu A/\mu m}$. Inclusion of lead scattering is, therefore,
crucial for predicting the performances of the 3HJ devices. The relatively small reduction of $I_{\rm{ON}}$ of the HJ TFETs is due to the band tail in the source.
Note, that Ref.~\cite{Khayera2011} predicted a stronger band tail effect using a much lager scattering rate. Overall, the 3HJ TFETs still possess much
larger $I_{\rm{ON}}$ than the HJ TFETs under equal scattering conditions.
We note that the 3HJ TFETs are more sensitive than the HJ TFETs to the gate misalignment: simulations (for Lg=15nm) show that although the 3HJ TFET can tolerate a small amount (1nm) of gate underlap with the source, a 1nm gate overlap with the source leads to a 47\% decrease in $I_{\rm{ON}}$ and a 140\% increase in $I_{\rm{OFF}}$ for the 3HJ TFET, compared with a 18\% decrease in $I_{\rm{ON}}$ and a 99\% increase in $I_{\rm{OFF}}$ for the HJ TFET. Therefore, processes with gate self-alignment are likely necessary and will be subject of future experimental studies. Further, the 3HJ TFET performance is also sensitive to the variations of the quantum well lengths, since the quantum well length critically determines the resonant energy level. Simulations show that, for Lg=15nm, +1 (-1) monolayer variation of the GaSb layer length leads to a 5.1\% increase (11\% decrease) in $I_{\rm{ON}}$ and a 9.5\% increase (14\% decrease) in $I_{\rm{OFF}}$; +1 (-1) monolayer variation of the InAs layer length leads to a 5.4\% increase (19\% decrease) in $I_{\rm{ON}}$ and a 140\% increase (54\% decrease) in $I_{\rm{OFF}}$. Note that one monolayer thickness in the [110] orientation is about 0.216nm. Therefore, the device performance is more sensitive to the channel well length variation and it needs to be controlled with $\pm1$ monolayer accuracy. Such precision in growth is achievable using atomic layer epitaxy \cite{Bedair1993selective}.

\section{Discussions}
The implementation of this model in this study has two major simplifications. First, the mobility is constant in the source and drain leads. However, the electron density distribution, especially in the source lead, is non-uniform due to the quantum well structure. This implies that the scattering rate and carrier mobility should be spatially varying. Moreover, the scattering rate is constant below $E_V$ and above $E_C$. Given the complicated band diagram and LDOS, the scattering rate should also be energy and momentum dependent. Therefore, a more accurate approach would need a more sophisticated density-dependent mobility model, as well as LDOS dependent scattering rate. Second, the channel is assumed to be ballistic in this model. In practice, phonon assisted tunneling would also occur in the channel to form a third leakage path. To model such leakage, the ballistic NEGF approach in the channel needs to be upgraded to include electron-phonon scattering in the usual SCBA way. We note, however, that the scattering in the leads is much stronger than that in the channel due to the high carrier concentrations in the leads. This has been observed even when only the electron-phonon scattering was included~\cite{Long2016a}. Consequently, the presented model captures the dominant scattering leakage path.

\section{Conclusion}
We have discussed that the effect of scattering in the leads is critical for 3HJ TFETs, and, at the same time, is very challenging to model in the NEGF approach. A multiscale model, that solves quantum drift-diffusion equations in the thermalizing scattering leads and NEGF equations in the ballistic channel, is developed to assess the realistic device performances. This model captures the important scattering leakage current in the sub-threshold region and serial resistance above threshold. Simulation results show, that, although the scattering degrades the SS and reduces the turn-ON current, with aligned $I_{\rm{OFF}}$ the 3HJ TFETs still operate with $<$60mV/dec SS and have much larger $I_{\rm{ON}}$ than the HJ TFETs for both 15nm and 30nm channel lengths. The 3HJ TFETs are thus very promising in future fast low-power computing applications. The model developed in this work can be used to study the effects of carrier thermalization and serial resistance in a variety of high-current TFETs, especially those featuring a potential notch in the source~\cite{pala2015exploiting,Verreck2014,Jiang2015}.


%



%
%

\ifCLASSOPTIONcaptionsoff
  \newpage
\fi

\end{document}